\documentstyle[aps]{revtex}
\newcommand{\ds}{\displaystyle}
\newcommand{\dsf}{\ds\frac}

\newcommand{\beq}{\begin{equation}}
\newcommand{\eeq}{\end{equation}}

\large
\begin{document}
\large
\begin{center}
\Large\bf
Nonlinear Stationary Waves  with Transport Current in Superconductors
\vskip 0.1cm
{\normalsize\bf N.А.\,Taylanov}\\
\vskip 0.1cm
{\large\em Department of Theoretical Physics,\\
Institute of Applied Physics,\\
National University of Uzbekistan,\\
Vuzgorodok, 800174, Tashkent, Uzbekistan\\
e-mail: taylanov@iaph.tkt.uz}
\end{center}
\begin{center}
\bf Abstract
\end{center}

\begin{center}
\mbox{\parbox{14cm}{\small
        The profile of a nonlinear stationary thermomagnetic wave in the
resistive state of superconductors is studied at different transport currents. It is
proved that the thermomagnetic wave has an oscillating profile at relatively
high values of the transport current in the sample. A shock wave with a
monotonic structure corresponds to comparatively weak transport currents.
The wave propagation velocity and the wave front width in a superconductor
are estimated.
}}
\end{center}
\vskip 0.5cm
        The problem of thermal suppression of the superconducting state in
the presence of transport current is topical in the practical use of
superconducting magnetic systems (see[1]). External fluctuations of various
physical natures (thermal, magnetic, mechanical, etc.) can lead to the
formation of resistive-phase regions in a superconductor. The ohmic heating
of these regions under a transport current with density $j_{tr}$ can bring
about an increase in the temperature $T$ above the critical point $(T>T_c)$
and, hence, the emergence of a vortex electric field $\vec E$
in the sample. It was shown earlier [2] that, together with dispersive and
nonlinear effects, the dissipative processes induced under ohmic heating
in a superconductor lead to the formation of stable structures,
thermoelectromagnetic $\vec E$ and $\vec H$ waves, depending on the condition
at the sample surface.
        In the present work, we studied the qualitative pattern of the
dynamics of nonlinear thermomagnetic dissipative structures in the resistive
state of a superconductor. The dynamics of thermomagnetic waves, which move
at a constant velocity $v$ along the $x$ axis in a superconductor, is
described in terms of self-simulated variables  $\xi(x,t)= x - vt$
by the nonlinear heat conduction equation [2-4]
\beq
-\nu v\dsf{dT}{d\xi}=\dsf{d}{d\xi}\left[\kappa\dsf{dT}{d\xi}\right]+jE,
\eeq
a set of Maxwell equations
\beq
\dsf{dE}{d\xi}=-\dsf{4\pi v}{c^2}j,
\eeq
\beq
E=\frac{v}{c}H
\eeq
and their related equation of the critical state
\beq
j=j_c(T,H)+j_{r}(E)+ j_{tr}
\eeq
Here, $\nu$ and $\kappa$ are the heat capacity and thermal conductivity
coefficients, respectively;
$j_{c}(T)=j_{0}-a(T-T_0)$ is the critical current density in the Bean
model $\left(\dsf{dj_{c}}{dH}=0\right)$ [5] (where $T_0$ and $T_c$ are the initial
and critical temperatures of the superconductor, respectively, and the
quantity $a=\left|\dsf{dj_c}{dT}\right|_{T=T_0}$ describes the thermally
activated weakening of the Abrikosov vortex pinning by lattice defects);
$j_{r}(E)$ is the resistive current density; and $j_{tr}$ is the transport
current density.
        In the weak-heating approximation $(T-T_0)<<(T_с-T_0)$
for the viscous flow region ($E>E_f$, where $E_f$ is the boundary of the
linear section in the current-voltage characteristics of hard
superconductors [6]), the dependence of the resistive current density $j_r$
on the electric field $E$ is linear; i.e., $j_r\approx \sigma_f E$
($\sigma_f$ is the effective conductivity). In weak fields $(E<E_f)$
the dependence $j_{r}(E)$ is essentially nonlinear and is most likely
determined by the thermally activated motion of the magnetic flux
(flux creep [6]). Here, we will consider a perturbation with a sufficiently
high amplitude $(E>E_{f})$ and use the linear dependence $j_{r}(E)$.
        The appropriate thermal and elecrtodynamic boundary conditions
for Eqs. (1)-(4) have the form
\beq
\begin{array}{l}
T(\xi\rightarrow+\infty)=T_0, \dsf{dT}{d\xi}(x\rightarrow-\infty)=0,\\
\quad\\
E(\xi\rightarrow+\infty)=0,   E(\xi\rightarrow-\infty)=E_e,\\
\end{array}
\eeq
where $E_e$ is the external electric field.

        By solving the set of Eqs. (1)-(4) with boundary conditions (5),
we obtain the following equation for the ditribution of the nonlinear
$E$ wave:
\beq
\dsf{d^3 E}{dz^3}+\beta(1+\tau)\dsf{d^2 E}{dz^2}+\beta^2\tau
\left[\dsf{dE}{dz}+\dsf{E_j}{E_\kappa}E\right]=0.
\eeq
Here, $z=\dsf{\xi}{L}$,$\beta=\dsf{vt_\kappa}{L}$,$E_j=\dsf{j_{tr}}{\sigma_f}$,
$\tau=\dsf{4\pi\kappa\sigma_f}{c^2\nu}$,$E_{\kappa}=\dsf{\kappa}{aL^2}$,
$L=\dsf{cH_e}{4\pi j_0}$ is the depth of the magnetic field penetration
into the bulk of the superconductor, $t_{\kappa}=\dsf{\nu L^2}{\kappa}$
is the thermal time of the problem, and $H_e$ is the external magnetic field.

        By analyzing the roots of the characteritic equation [7]
\beq
\lambda(\lambda+\beta)(\lambda+\beta\tau)=-\beta^2\tau\dsf{E_j}{E_\kappa},
\eeq
we can easily verify that the problem considered has solution that
oscillate about the origin of the coordinates $(E_0=0)$
of the phase diagram with decreasing and increasing amplitudes. The
oscillations appear at $E_j>\dsf{\beta\tau}{2(1+\tau)} E_\kappa$,
attenuate at $\beta(1+\tau) E_\kappa >E_j>\dsf{\beta\tau}{2(1+\tau)}
E_\kappa$, and increase at $\beta(1+\tau) E_\kappa< E_j$.
This means that, at a sufficiently high transport current  $j_{tr}$,
the roots of the characteristic equation (7) are complex, the integral
curves are helical, and the origin $E_0=0$ is a singular point (focus).
If the coefficients $\beta$ and $\tau$ are assumed to be universally
positive, the representative point approaches the origin of coordinates,
which is a stable equilibrium position.
In the opposite case, when $\beta(1+\tau)E_\kappa<E_j$,
the focus becomes unstable and the representative point goes to infinity.
Note that the final value of the transport current $j_{tr}$
leads to a displacement of singular points and, correspondingly, of
the boundary conditions.

        At sufficiently weak transport currents, when the condition
  $\dsf{dE}{dz}>>\beta^2\tau E_{j}$
is satisfied (high-amplitude waves), Eq. (6) is an integrable equation
and the corresponding model can be considered a damped linear oscillator
owing to the friction force, that is,
\beq
F_t=-\beta(1+\tau)\dsf{dE}{dz}\,,
\eeq
where $z$  is an analogue of time and $E$ is the coordinate of the
"material" point. The potential well equation has the form
\beq
U(E)=-\dsf{E^3}{6E_\kappa}+\beta^2\tau\dsf{E^2}{2}.
\eeq

        Analysis of the phase plane $(\dsf{dE}{dz},E)$
described by Eq. (6) shows that it has two equilibrium points: $E_0$ is
a stable node and $E_1=2\beta^2\tau E_\kappa$ is a saddle.
The two equilibrium states $(E_0,E_1)$ are separated by the separatrix $AB$
in the phase plane (Fig.1). The material point is located at the point
$E_1$ at $z\rightarrow \infty$  and goes over into the point $E_0$ at
$z\rightarrow -\infty$.
The change-over from one equilibrium state to other occurs only
monotonically.

        Within the approximation   $\tau<<1$, the solution of Eq. (6)
is represented as
\beq
E(z)=\dsf{E_1}{2}\left[1-th(\dsf{\beta}{2} (z-z_0)\right]
\eeq
        The condition  $\tau=\dsf{D_t}{D_m}<<1$ means that the magnetic
flux is redistributed much faster than the heat is transferred. Here,
$D_t=\dsf{\kappa}{\nu}$ and $D_m=\dsf{c^2}{4\pi \sigma_f}$ are the
thermal and magnetic diffusion coefficients, respectivily [6].
Hence, the spatial scale $L_E$ for the magnetic flux penetration is
substantially greater than the corresponding thermal scale $L_T$.
Therefore, the spatial derivatives $\dsf{d^n E}{dz^n}$
will contain a small parameter $\left[\dsf{L_T}{L_E}\right]<<1$.
We can easily check the validity of this approximation by the direct
differentiation of Eq. (6), that is,
\beq
{\dsf{d^2 E}{dz^2}}\left(\beta{\dsf{dE}{dz}}\right)^{-1}=\tau<<1
\eeq

        The maximum error of this approximation is of the order
$\dsf{\tau}{(1+\tau)^2}$ .
For example, the error is equal to 25 
small in the limiting case $\tau\rightarrow 0$. Relationship (10)
describes the profile of a shock thermomagnetic wave propogating into
the bulk of the superconductor.
The wave structure is shown in Fig.2. At the boundary condition
$(E(z\rightarrow-\infty)=E_e)$, we can easily determine the velocity
$v_E$ of the wave with an amplitude $E_e$ in the form
\beq
v_E=\dsf{L}{t_\kappa}\left[\dsf{E_e}{2\tau E_\kappa}\right]^{1/2},
\eeq
and the wave front width $\delta z$
\beq
\delta z=16\dsf{1+\tau}{\tau^{1/2}}\left[\dsf{E_\kappa}{E_e}\right]^{1/2}.
\eeq

        The numerical estimates yield the values
$v_E=1\div 10^2 \dsf{см}{сек}$ and $\delta z=10^{-1}\div 10^{-2}$
for $\tau=1$.

        In conclusion, we note that the results obtained make it possible
to describe the nonlinear stage of evolution of the thermomagnetic
instability in the resistive state of superconductors. In the initial stage,
we observe an exponential increase in the perturbations of $T$, $E$, and $H$
with the increment $\lambda$ determined from the linear theory (see [8]).
The linear stage of the instability evolution lasts for the period
$t_j=\dsf{t_\kappa}{\lambda}$.
In the latter stage of the instability evolution, the stationary
$\vec E$ or $\vec H$ wave (depending on whether the $\vec E$ or $\vec H$
quantity is fixed on the sample surface during the wave motion) propagates
into the bulk of the sample. The velocity of the wave is determined by
formula (12),
and the time of its motion in the sample with thickness $2l$ is given by
$$
\Delta t=\dsf{l}{v_E}
$$
Hence, it follows that the transition from the superconducting to the
normal state can occur through the propagation of a stationary thermomagnetic
wave whose structure essentially depends on dispersive and dissipative
effects.

\begin{center}
REFERENCES
\end{center}
\begin{enumerate}
\item
V.R.Romanovskii, Dokl. Akad. Nauk 365, (1) 44(1999)
[Dokl.Phus. 44, 137 (1999)].
\item
I.L.Maksimov, Yu.N.Mastakov, and N.A.Taylanov.
Fiz. Tverd. Tela (Leningrad) 28 (8), 2323(1986)
[Sov. Phys. Solid State 28, 1300 (1986)].
\item
N.A. Taylanov, Metallofizika (Kiev) 13 (9), 713(1991).
\item
N.A.Taylanov Superconduct. Science and Technology, 14, 326 (2001).
\item
Bean C.P. Phys. Rev. Lett. 8,6. 250 (1962).
\item
R.G. Mints and A.L. Rakhmanov, Instability in Superconductors
(Nauka, Moscow, 1984).
\item
V.I. Karpman, Non-linear Waves in Dispersive Media
(Nauka, Moscow, 1973; Pergamon, Oxford, 1975).
\item
V.A. Al'tov, V.B. Zenkevich, M.G. Kremlev, and V.V. Sychev, Stabilization
of Superconducting Magnetic Systems (Energoatomizdat, Moscow, 1984).
\end{enumerate}
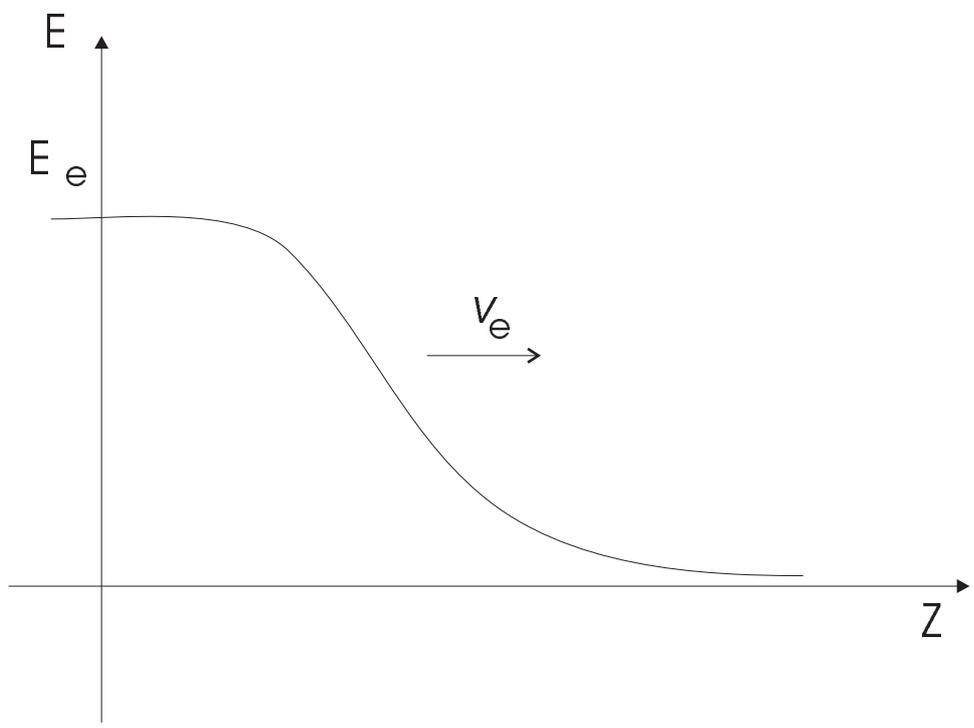
\end{document}